\documentclass[twocolappendix]{emulateapj}
\let\oldvec\vec

\usepackage{amsmath,amssymb,amstext}
\usepackage{mathtools}
\usepackage{gensymb}
\usepackage[usenames,dvipsnames]{color}
\usepackage{tabularx}

\newcolumntype{Y}{>{\centering\arraybackslash}X}
\usepackage{threeparttable}
\usepackage{ulem}
\usepackage[colorlinks,urlcolor=blue,citecolor=blue,linkcolor=blue,pdfusetitle]{hyperref}

\usepackage{color}

\newcommand{\alr}[1]{{\textcolor{teal}{(ALR: {#1})}}}

\usepackage{comment}
\usepackage{todonotes}
\shorttitle{Circumbinary Disks}
\shortauthors{Duffell et al.}

\begin{document}

\title{Circumbinary Disks: Accretion and Torque as a Function of Mass Ratio and Disk Viscosity}

\def\harv{1}
\def\colum{2}
\def\nyu{3}
\def\zrake{4}

\author{Paul C. Duffell\altaffilmark{\harv}, Daniel D'Orazio\altaffilmark{\harv}, Andrea Derdzinski\altaffilmark{\colum}, Zoltan Haiman\altaffilmark{\colum}, Andrew MacFadyen\altaffilmark{\nyu}, Anna L. Rosen\altaffilmark{\harv}, and Jonathan Zrake\altaffilmark{\colum}}

\altaffiltext{\harv}{Center for Astrophysics $|$ Harvard \& Smithsonian, 60 Garden Street, Cambridge MA 02138}
\altaffiltext{\colum}{Department of Astronomy, Columbia University, New York, NY 10027}
\altaffiltext{\nyu}{Center for Cosmology and Particle Physics, New York University, New York, NY 10003}
\email{paul.duffell@cfa.harvard.edu}

\begin{abstract}
Using numerical hydrodynamics calculations and a novel method for densely sampling parameter space, we measure the accretion and torque on a binary system from a circumbinary disk.  In agreement with some earlier studies, we find that the net torque on the binary is positive for mass ratios close to unity, and that accretion always drives the binary towards equal mass.  Accretion variability depends sensitively on the numerical sink prescription, but the torque and relative accretion onto each component do not depend on the sink timescale.  Positive torque and highly variable accretion occurs only for mass ratios greater than around $0.05$.  This means that for mass ratios below $0.05$, the binary would migrate inward until the secondary accreted sufficient mass, after which it would execute a U-turn and migrate outward.  We explore a range of viscosities, from $\alpha = 0.03$ to $\alpha = 0.15$, and find that this outward torque is proportional to the viscous torque, so that torque per unit accreted mass is independent of $\alpha$.  Dependence of accretion and torque on mass ratio is explored in detail, densely sampling mass ratios between $0.01$ and unity.  For mass ratio $q > 0.2$, accretion variability is found to exhibit a distinct sawtooth pattern, typically with a five-orbit cycle that provides a ``smoking gun" prediction for variable quasars observed over long periods, as a potential means to confirm the presence of a binary.

\end{abstract}


\section{Introduction} \label{sec:intro}

A gaseous disk in orbit around a binary system is a physical scenario relevant to a range of questions in astrophysics \citep{1994ApJ...421..651A}.  During the formation of a stellar binary, a protostellar disk is typically present, which can affect the eventual orbital parameters and mass ratio of the binary \citep{2000MNRAS.314...33B}.  It is thought that disk accretion onto the binary could be responsible for the observed population of ``twin" binaries with mass ratios $q \gtrsim 0.95$ \citep{2019arXiv190610128E}.  Binary black holes may also encounter a circumbinary disk in their lifetime, especially a supermassive black hole binary at the center of a galaxy following a merger.  In this case, torques from an accretion disk may play a role in bringing the binary close enough to merge \citep{2009ApJ...700.1952H}.  The effects of the disk on the binary may be observable in the gravitational wave background as viewed by pulsar timing \citep{2011MNRAS.411.1467K, 2019MNRAS.485.1579K}.  Disk torques can also potentially provide a correction to the inspiral, affecting the gravitational wave signal observed by LISA \citep{2019MNRAS.486.2754D}.

Understanding how the disk affects the binary requires the calculation of torques and accretion rates onto each binary component as a function of disk viscosity ($\alpha$), aspect ratio ($h/r$), and the binary mass ratio ($q$).  Such a calculation is nontrivial, as it requires the evolution of a highly nonlinear gas dynamics system over thousands of binary orbits.

Two of the most important recent results from such calculations are the finding of \cite{2014ApJ...783..134F} that accretion always drives the binary toward equal mass, and the highly counter-intuitive result that the disk torques can be \textit{positive}, meaning that the binary can be driven outward rather than inward \citep{2017MNRAS.469.4258T, 2019ApJ...871...84M, 2019ApJ...875...66M, 2019arXiv191004763M}.

The present numerical study takes on a more thorough search of parameter space, to determine just how robust these results are to variations in mass ratio and viscosity\footnote{Dependence on the disk Mach number will not be tested in this study, though this is another parameter whose dependence should be measured, because real black hole accretion disks are typically much thinner than what is modeled numerically.  This dependence will be checked in a future study \citep[][has previously studied dependence of this system on Mach number]{2016MNRAS.460.1243R}.}.  To do so, we employ a novel technique whereby parameter space is ``scanned" continuously by slowly varying the binary mass ratio with time.  This makes it possible to thoroughly explore parameter space in only a handful of numerical runs.

The numerical method is described in Section \ref{sec:numerics}.  Results of these calculations are detailed in Section \ref{sec:results}, including the fiducial disk model and an investigation of dependence on disk parameters.  All of the results will be discussed in Section \ref{sec:disc}, including implications for stellar and black hole binaries.

\begin{figure}
\includegraphics[width=3.5 in]{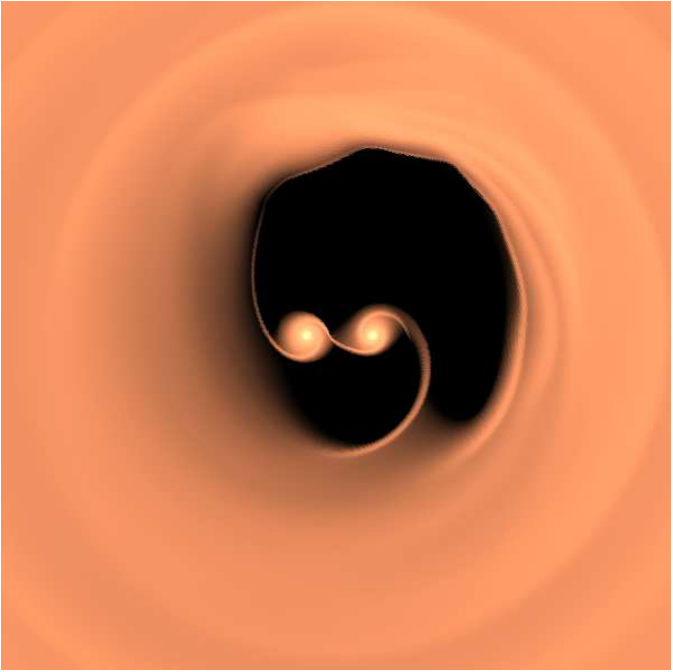}
\caption{Logarithm of density in a circumbinary disk with mass ratio $q = 0.8$ after 1000 orbits.  Depicted surface density ranges from $\Sigma/\Sigma_0 = 0.01$ to $100$.} 
\label{fig:snapshot}
\end{figure}

\section{Methods} \label{sec:numerics}

This study uses the moving-mesh code \texttt{Disco} \citep{2016ApJS..226....2D} to numerically integrate the equations of isothermal hydrodynamics in 2D:

\begin{equation}
\partial_t \Sigma + \nabla \cdot (\Sigma \vec v) = S_{\Sigma}
\end{equation}
\begin{equation}
\partial_t ( \Sigma v_j ) + \nabla \cdot ( \Sigma \vec v v_j + P \hat x_j - \nu \Sigma \vec \nabla v_j ) = - \Sigma \vec \nabla \phi + S_{\Sigma} \vec v
\end{equation}
\begin{equation}
P = c^2 \Sigma
\end{equation}
where $\Sigma$ is the surface density, $P$ denotes pressure, $\vec v$ is the velocity, $\nu$ is the kinematic viscosity, $c$ is the sound speed, and $\phi$ is the binary system's gravitational potential.  The sink term $S_{\Sigma}$ is used for accretion onto each body and will be described later.

The sound speed is computed locally based on the distance to each binary component.  Specifically, it is computed using the local gravitational potential:

\begin{equation}
    c = \sqrt{\phi}/\mathcal{M}.
\end{equation}
\noindent
Kinematic viscosity $\nu$ is taken to be a constant, but it is parameterized in terms of the quantity $\alpha = \mathcal{M}^2 \nu / (a^2 \Omega_B)$.  The Mach number $\mathcal{M}$ is taken to be $10$ for all calculations in this study, implying a disk aspect ratio of $h/r = 1/\mathcal{M} = 0.1$.  The gravitational potential is cut off by a smoothing length:

\begin{equation}
    \phi_i = \frac{G M_i}{\sqrt{ | \vec x - \vec x_i |^2 + \epsilon^2 }}.
\end{equation}
\noindent
This length scale is chosen to be $\epsilon = \frac12 a / \mathcal{M} = 0.05 a$ where $a$ is the semimajor axis.

The initial disk model is taken to be a uniform surface density disk, with $\Sigma = \Sigma_0$.  The orbital velocity $\Omega$ is initially given by the Keplerian orbital velocity, except within $r<a$ 
where it is given by rigid rotation with the binary orbital frequency\footnote{The initial velocity within $r<a$ is essentially irrelevant, as this information is erased within a few orbits as the cavity is cleared.}.  Finally, $v_r$ is initialized as $-\frac32 \nu/r$.

The standard resolution is given by $512$ radial zones logarithmically spaced with $\Delta r/a = 0.011$ at the binary orbital radius.  The outer radius is set at $r = 30 a$.  This large outer boundary is chosen in order to minimize the effects of long-distance torques from the binary, so that the accretion rate will solely be determined by the outer boundary, with no influence by the binary.

Mass is removed close to each binary component via the following sink prescription.  An additional evolution term for the surface density is

\begin{equation}
    S_\Sigma = - \gamma \Sigma \left( \rm exp \left( -\frac{ | \oldvec x - \oldvec x_1 |^4}{\epsilon^4} \right) + \rm exp \left( -\frac{ | \oldvec x - \oldvec x_2 |^4}{\epsilon^4} \right) \right),
\end{equation}
\noindent
where $\gamma$ is a numerical parameter that is varied to determine whether the sink prescription affects the torque on the binary.  For our fiducial run, we use our fastest sink rate, $\gamma = 3 \Omega_B$.  We find that the accretion rate we measure onto the binary is independent of $\gamma$ and consistent with the accretion rate at infinity, so long as $\gamma > 0.01 \Omega_B$.  We choose the length scale for the sink to be the same as the smoothing length $\epsilon$, as this is the length scale below which we do not trust the solution's accuracy.

The treatment of binary potential is the same as described in the \texttt{Disco} code paper \citep{2016ApJS..226....2D}, except that in this study, a novel approach is employed: rather than choosing a single mass ratio and running the calculation until the disk is in a stationary quasi-steady-state, the mass ratio is very slowly decreased from $q = 1$ to $q = .01$, at a rate slow enough such that at any given time the binary can be considered as interacting with a steady-state\footnote{The disk is not actually in steady-state; it can be highly variable, as we show later.  What is meant by "steady state" here is that there is no secular evolution to the disk.} disk:

\begin{equation}
    q(t) = \frac{1}{1 + (t/\tau)^2},
\end{equation}
with $\tau = 2000$ orbits for the fiducial run; generally, $\tau$ is chosen to be a fixed number of viscous times.  As the binary is evolved, its trajectory is adjusted to follow a circular, two-body orbit for a binary with fixed separation $a$ and instantaneous mass ratio $q$ whose center of mass is at the origin.  The calculations were run for mass ratio as small as $q = 0.01$, at which point the tidal truncation radius becomes comparable to the sink radius \citep{2014ApJ...785..115R}.

The benefit of this method is that one can measure the torque and accretion continuously as a function of mass ratio, instead of performing many separate numerical calculations to sample a discrete number of points in parameter space.  The main diagnostics being measured are the total torque on the binary and the accretion rates onto each binary component.

The torque experienced by the binary can be broken into two components: one due to gravitational force from fluid elements in the disk, and the other by direct accretion of mass:

\begin{equation}
    \dot J_{\rm bin} = T_{\rm grav} + \dot J_{\rm acc}
\end{equation}

The accretion term $\dot J_{\rm acc}$ can be further broken into pieces; if the accreted gas has zero momentum in the rest frame of the point mass, the point mass increases its momentum simply due to its increase in mass.  If one knows the relative accretion rate onto each component, one can analytically calculate this term by $\dot M_1 j_1 + \dot M_2 j_2$, where $j_1$ and $j_2$ are the specific angular momentum of each point mass.  However, if the accreted mass has nonzero momentum in the rest frame of the point mass, there is a residual term determined by the momentum accreted in the rest frame of the point mass:

\begin{equation}
    \dot J_{\rm acc} = \dot M_1 j_1 + \dot M_2 j_2 + \Delta \dot J_{\rm acc}.
    \label{eqn:jdot}
\end{equation}

In all cases studied, we found the last term to be negligible when compared to the first two.  That is, the mass accreted is moving at very nearly the same velocity as the corresponding point mass.  This means it is possible to use the first two terms to calculate the accretion torque exactly, given the relative accretion rates $\dot M_1$ and $\dot M_2$.

The torque reported in this study is only the gravitational term $T_{\rm grav}$, as we have found $\Delta \dot J_{\rm acc}$ to be negligible, and $\dot J_{\rm acc}$ is therefore accurately given by the first two terms in (\ref{eqn:jdot}).  The gravitational force we calculate is smoothed over 30 orbits using a gaussian convolution.  Note there is a difference between the way we report our torque and the way reported by \cite{2019arXiv191004763M}, in that the Mu\~noz study combines the gravitational torque with the accretion term ($\dot J_{\rm acc} = 0.25 \dot M a^2 \Omega_B$ for an equal mass binary) accounting for the angular momentum accreted at the binary's specific angular momentum.

\begin{figure}
\includegraphics[width=3.5 in]{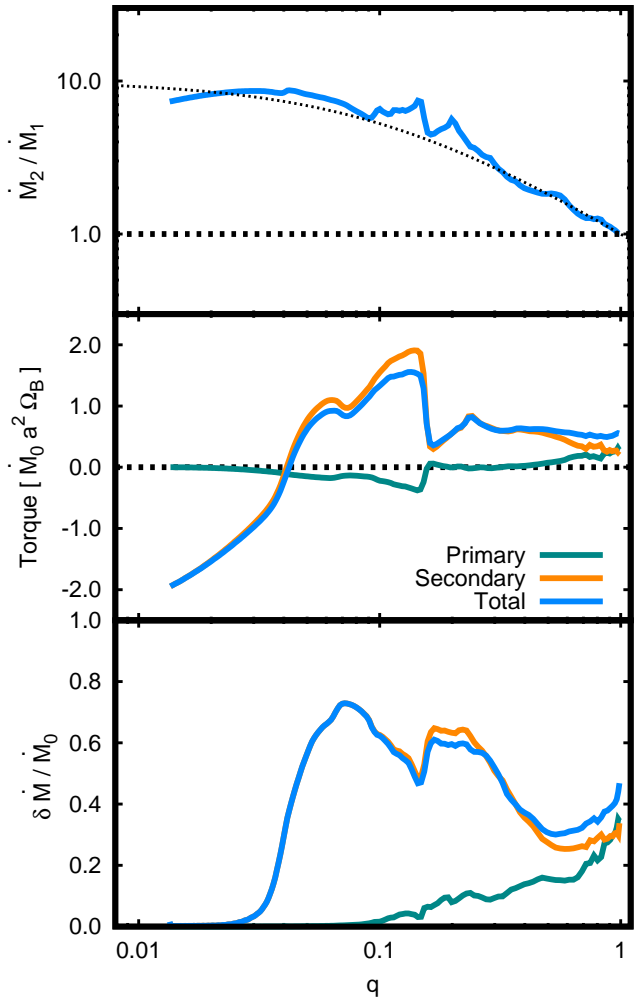}
\caption{Measured quantities as a function of mass ratio for the fiducial disk.  Top panel: relative accretion onto the secondary ($\dot M_2 / \dot M_1$).  Center panel: Torque.  Bottom panel: Accretion Variability.}
\label{fig:over}
\end{figure}

\section{Results}
\label{sec:results}

\subsection{Overview}

A snapshot of an example calculation is shown in Figure \ref{fig:snapshot}.  We begin with a basic overview of dependence of the system on mass ratio for the fiducial disk (Figure \ref{fig:over}).  The fiducial disk has $\alpha = 0.1$, $h/r = 0.1$, implying a viscous timescale of $a^2/(3 \pi \nu) \approx$ 100 orbits.  For all disk parameters and choices of sink prescriptions, we recover that the total accretion rate $\dot M = \dot M_1 + \dot M_2$ onto the binary is consistent (to within $20\%$) with the analytical accretion rate set at the outer boundary, $\dot M_0 = 3 \pi \nu \Sigma$.  We find that in the range of binary parameters explored, accretion always favors the secondary, as shown in the top panel of Figure \ref{fig:over}.  When $q=1$, $\dot M_1 = \dot M_2$ by the symmetry of the system, but as $q$ is reduced the ratio $\dot M_2 / \dot M_1$ becomes greater than unity, rising to $\sim 10$, implying that the secondary eats about $90\%$ of the gas for mass ratios $\sim 0.05$.  Our calculations only considered mass ratios $> 0.01$, but calculations in the context of planetary accretion have claimed $90\%$ accretion onto the secondary for $q = 10^{-3}$ \citep[Jupiter's mass;][]{2006ApJ...641..526L}.  We find this curve is reasonably well-fit by the formula

\begin{equation}
    \frac{\dot M_2}{\dot M_1} = \frac{1}{0.1 + 0.9 q},
    \label{eqn:lambda}
\end{equation}
as denoted by the dotted line in the figure.  This formula provides an update to the accretion formula of \cite{2019MNRAS.485.1579K} which was found by a fit to the data points in the study of \cite{2014ApJ...783..134F}.

Torque has a richer $q$ dependence, as shown in the center panel of Figure \ref{fig:over}.  At $q = 1$, we find the torque is positive and equal to $0.6 \dot M a^2 \Omega_B$, broadly consistent with other studies that find positive torque on the binary; \cite{2019ApJ...871...84M, 2019arXiv191004763M} reported a normalization of $0.43$, and \cite{2019ApJ...875...66M} report $0.47$\footnote{These studies actually report a value of 0.68 and 0.72, respectively, but they also include accretion torque $\dot J_{\rm acc}$ in this calculation, which equals $0.25 \dot M a^2 \Omega_B$ for an equal-mass binary.  The values of $0.43$ and $0.47$ are quoted to directly compare with our gravitational torque $T_{\rm grav}$.}.  As the mass ratio is reduced, we find two transitions.  First, around $q \approx 0.15$, there is a jump in the torque by a factor of a few.  At this point, almost all the torque is due to interaction with the secondary.  Around $q \approx 0.05$, the torque flips sign, and inward migration is recovered for smaller mass ratios, with the magnitude of the torque being roughly equal to the viscous torque $\dot M a^2 \Omega_B$.

Variability also depends on the binary mass ratio.  As the mass ratio is reduced from $q=1$, the variability becomes dominated by the secondary (because the overall accretion is dominated by the secondary).  This variability remains strong until $q < 0.05$, where variability diminishes and accretion becomes steady.  This was also found previously by \cite{2016MNRAS.459.2379D} in a systematic study of this transition point. 

\begin{figure}
\includegraphics[width=3.5 in]{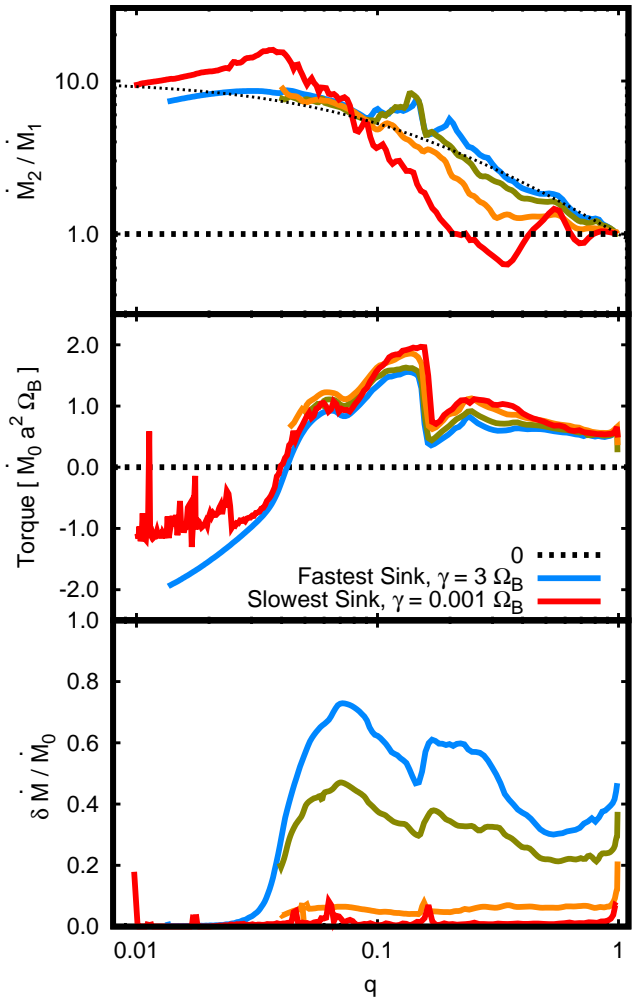}
\caption{Dependence on the sink prescription.  Red, orange, green, and blue show $\gamma/\Omega_B = 0.001$, $0.01$, $1$, and $3$ respectively.  The torque is notably independent of the sink timescale, given by the inverse of $\gamma$.  The accretion ratio is only affected when $\gamma$ is very slow, $< 0.01 \Omega_B$.  The viscous rate at the sink radius would correspond to $\gamma = \nu/\epsilon^2 = 0.4 \Omega_B$, which is closest to the green curve.} 
\label{fig:sink}
\end{figure}

\subsection{Dependence on imposed timescales}

Before exploring dependence on viscosity, we first determine how these measured quantities depend on the time-variation of $q$ and on the sink prescription.  The treatment of the gas very close to each point mass is nontrivial; it is unclear what the ``correct" prescription should be for the sink, and any quantity that depends on such numerical choices should be treated with care.

Fortunately, two out of three of our measured quantities appear to be roughly independent of the sink timescale.  Figure \ref{fig:sink} shows dependence of (time-averaged) torque (center panel) and relative accretion (top panel) on the sink timescale.  Faster sinks appear to lead to slightly more accretion onto the secondary, but the distinction is not significant.  Importantly, the torque also appears to not depend on this numerical choice, providing more evidence that these positive torques are not a numerical artifact.  This is in contrast with the results of \cite{2017MNRAS.469.4258T} which found strong dependence of the torque on the sink rate.  The reason for this discrepancy is unknown.  Our findings are consistent with recent work by \cite{2019ApJ...875...66M}, who also demonstrated the independence of torques on the sink timescale.

After the submission of the original version of our manuscript, \cite{2019arXiv191004763M} also presented a study covering a handful of mass ratios and viscosities, but they did not see a sharp jump in torque that we find around q = 0.15.  It is possible this difference is due to the size of their parameter space, but future studies should explore whether this sharp jump is a genuine physical effect.  Additionally, \cite{2019arXiv191004763M} did not see any transition in the sign of the torque, though they argue the binary may migrate inwards for sufficiently low mass ratio even with a positive torque.

However, variability is strongly dependent on the sink timescale \citep[this was also pointed out by][]{2014ApJ...783..134F}.  For sufficiently fast sink timescales, variability is given by how quickly the minidisks are fed, but for slow sinks, the minidisks can ``buffer" the accretion, smoothing out these variations.  In the limit of a very slow sink, there is no accretion variability.  Because the accretion variability is potentially an observable physical effect, one cannot ignore the numerical choice of sink prescription entirely.  In this study, results will be quoted in terms of our fastest sink ($\gamma = 3 \Omega_B)$, with the understanding that the time variability of a realistic system could be less extreme than what is measured here.

\begin{figure}
\includegraphics[width=3.5 in]{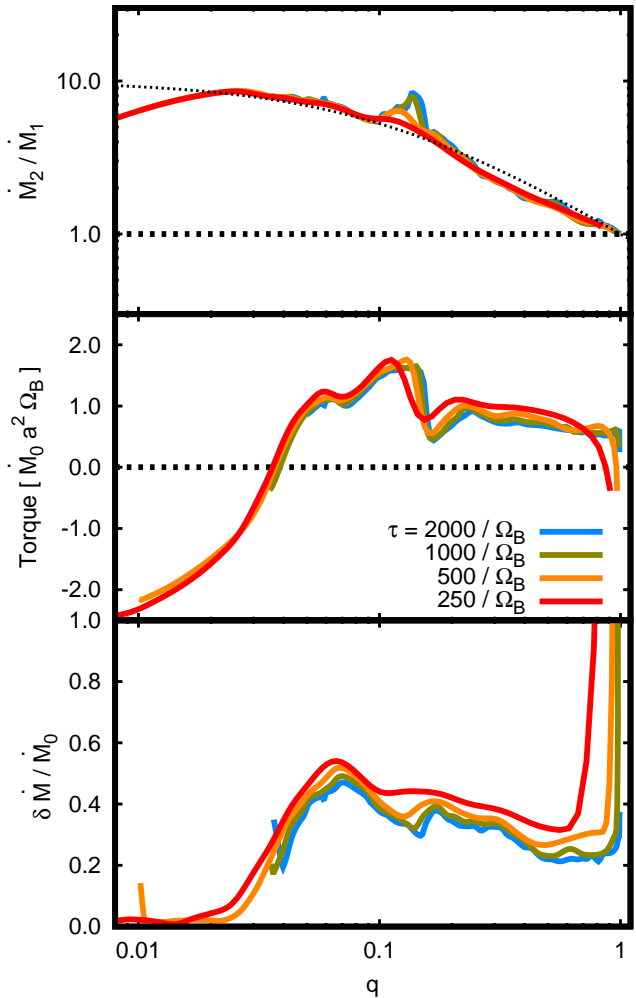}
\caption{Varying the timescale $\tau$ over which the mass ratio is changed.  The fiducial choice $\nu \tau/a^2 = 2$ used in this study appears to be significantly slower than necessary.  The biggest difference between different values of $\tau$ is the start-up transient near $q=1$.  It takes some time for the instability to develop and for the cavity to become lopsided, after which even our fastest-varying mass ratios would have recovered the result to adequate precision.} 
\label{fig:times}
\end{figure}

Additionally, several calculations were carried out with different choices of the parameter $\tau$, the timescale over which the binary mass ratio is varied (Figure \ref{fig:times}).  The qualitative impact of the changing $q$ is a ``smearing-out" of these curves.  The only real difference in the torque for shorter $\tau$ is at the earliest times, at largest $q$.  This ``spike" at $q \approx 1$ most likely occurs because the numerical calculation is initiated with $q=1$, and therefore the net torque near $q=1$ could be contaminated by a start-up transient.  By using the value of $\tau = 2000$ ($\nu \tau / a^2 = 2$), we can eliminate the effects of this start-up transient phenomenon.  This demonstrates the importance of running the calculation for a sufficiently long time in order to correctly recover the positive torques observed.

\subsection{Dependence on Viscosity}

\begin{figure}
\includegraphics[width=3.5 in]{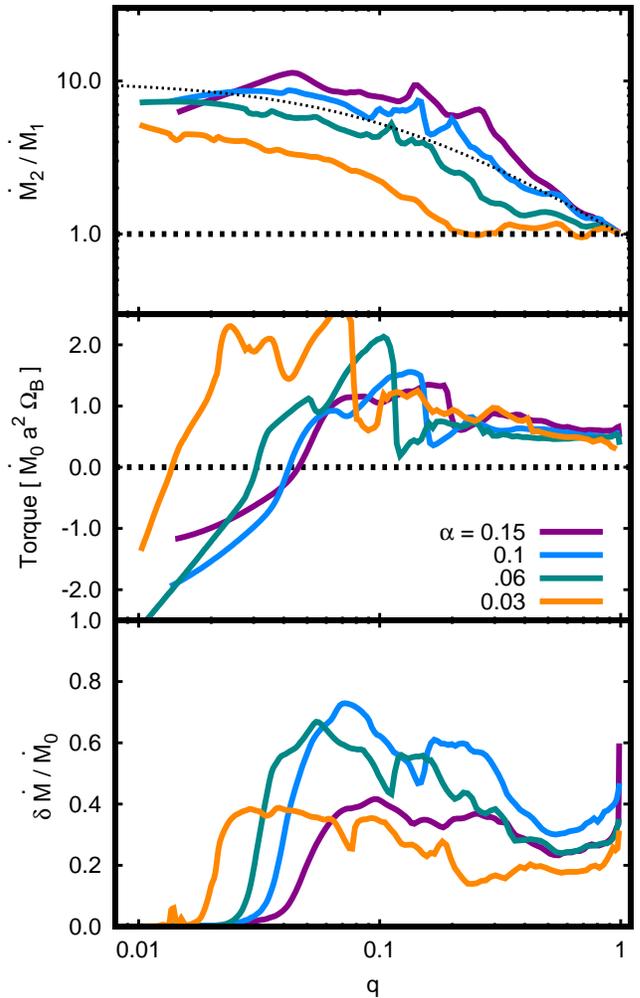}
\caption{Accretion and Torque as a function of mass ratio and disk viscosity.  The positive torque found for $q > 0.1$ appears to be consistently positive and close to $0.6 \dot M a^2 \Omega_B$, for all values of $\alpha$ considered.  This is consistent with the findings of \cite{2019ApJ...871...84M}, but this now appears to be a viscosity-independent statement, which also applies over a large range of mass ratios.  The transition to a highly variable disk at $q \approx 0.05$ appears to be a viscosity-dependent transition.  It appears to roughly coincide with the sign-reversal of the torque.} 
\label{fig:alphas}
\end{figure}

Most numerical studies of binaries use a rather large viscosity (either by using a large $\alpha$ or a large $h/r$), because it is necessary to wait for a viscous time before one can ascertain the time-averaged torques.  Here we investigate how the torques and accretion depend on viscosity.

We explore a range of viscosities from $\alpha = 0.03$ to $0.15$, keeping $h/r$ fixed at $0.1$.  Lower viscosity than $\alpha = 0.01$ was not attempted, because of the prohibitively long timescales that one must wait in order to attain a steady-state.  The fiducial run used $\nu = 10^{-3} a^2 \Omega_B$, with $\tau = 2000$ orbits.  As viscosity is varied, the timescale $\tau$ is also varied such that $\nu \tau / a^2 = 4 \pi \approx 13$ is kept fixed.  This gives the disk time to attain steady-state as the mass ratio varies.

For a wide range of values of $q$ and $\alpha$ the normalized torque is independent of viscosity, except in the sense that it is proportional to the accretion rate; $T / (\dot M a^2 \Omega_B)$ is a fixed constant for most $q \gtrsim 0.2$.





The overall conclusion is that the torque for $q \gtrsim 0.05$ is robustly positive and approximately equal to $0.6 \dot M a^2 \Omega_B$, independent of viscosity, mass ratio, or sink timescale.  The question of whether equal-mass binaries are ever pushed inward by a disk is still an open one, though observations of stellar binaries suggest that inward migration occurs in many cases \citep{2017ApJS..230...15M, 2019arXiv190610128E}.  Perhaps a binary with significant eccentricity will experience stronger inward torques by the disk; it has been shown that the disk can drive the binary to be eccentric, even in the regime of low mass ratio \citep{2001A&A...366..263P, 2013MNRAS.428.3072D}.  At high mass ratios, it is possible that the asymmetry of the cavity will lead to a strong driving of the binary towards higher eccentricity.  Some hints of this have already been shown in the handful of eccentric calculations by \cite{2019ApJ...871...84M}.  We will explore this possibility in a future study.

Relative accretion also depends on the viscosity.  For reduced $\alpha$, accretion is more evenly divided between the primary and the secondary.  This reduced value of $\dot M_2 / \dot M_1$ is still high enough that the binary trends toward equal mass, but not as quickly.

\subsection{Variability Timescales}

\begin{figure}
\includegraphics[width=3.5 in]{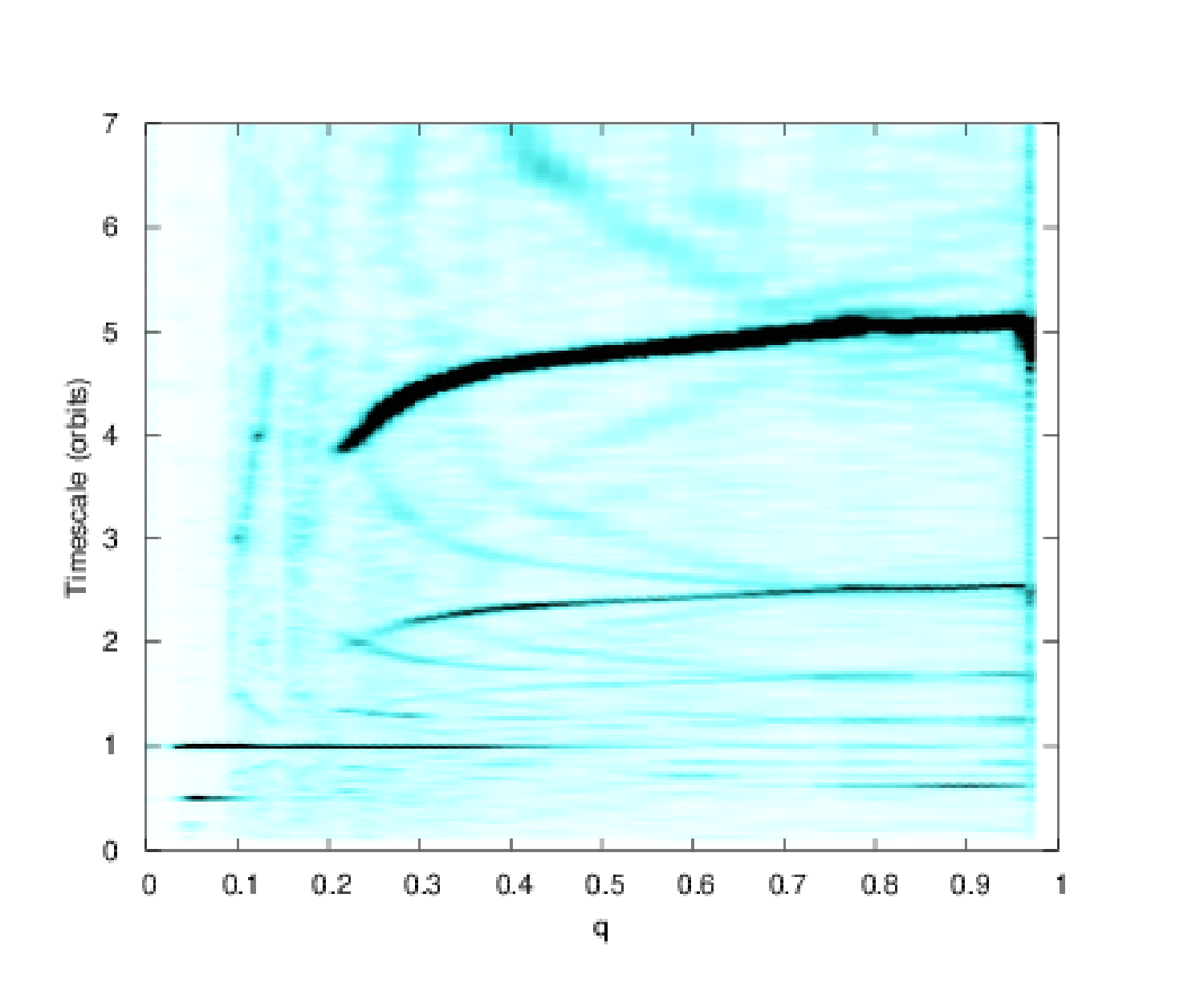}
\caption{2D Periodogram showing variability with respect to different timescales.  Longer timescale (4 to 5 orbit) variability occurs for $q > 0.2$.  $\delta \dot M$ is calculated at each frequency via equation (\ref{eqn:dmdot}).}
\label{fig:vari_times}
\end{figure}

\begin{figure}
\includegraphics[width=3.5 in]{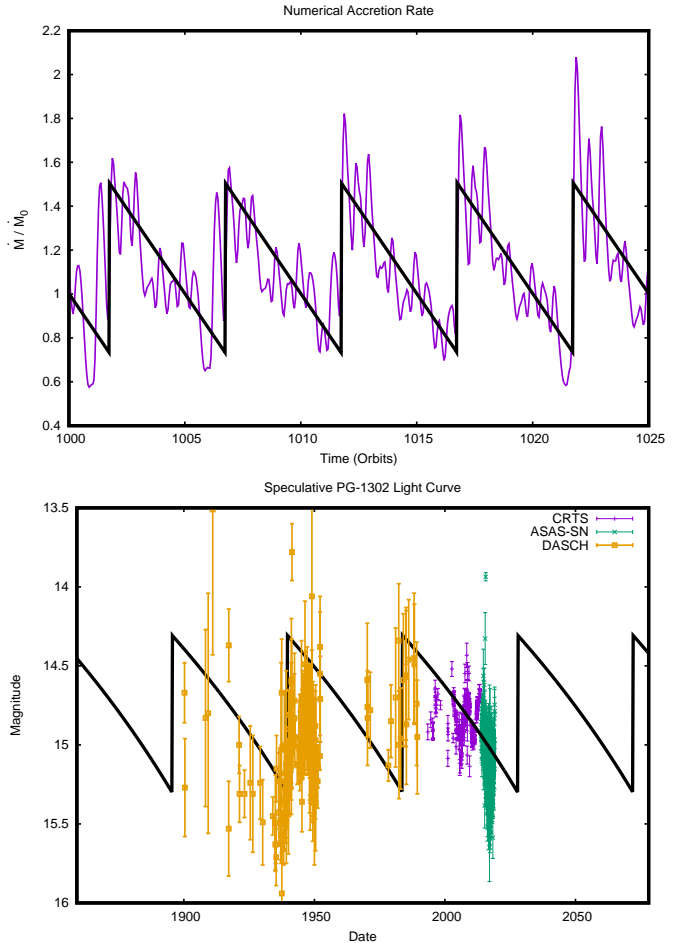}
\caption{Top Panel:  5-orbit variability roughly follows a sawtooth pattern in time as mass piles up at the cavity edge and falls onto the binary all at once every five orbits.  If this long-term variability is responsible for deviations from sinusoidal behavior in PG-1302, then we predict a sharp rise in the accretion rate at the end of the five-orbit cycle.  Bottom Panel:  A possible variability scenario for PG-1302, assuming the observed short-term variability is on the half-orbit timescale.  This scenario predicts a rapid rise, peaking in magnitude around the year 2027.  This is only one possible scenario, but if this rapid rise were observed, it would be strong evidence in favor of PG-1302 housing a binary.} 
\label{fig:saw}
\end{figure}

Figure \ref{fig:over} showed that the accretion is highly variable for all mass ratios greater than $q = 0.05$.  However, variability occurs on many different timescales, as shown by \cite{2014ApJ...783..134F}.  Figure \ref{fig:vari_times} shows the contribution to the variability on different timescales.

This point is typically demonstrated using a periodogram of the accretion rate at different specific mass ratios.  We choose to present the periodogram as a continuous function of $q$.  The accretion-rate variability amplitude at angular frequency $\omega$ is computed by evaluating the Fourier transform of the accretion rate at $\omega$, windowed over a short time interval to account for the time growing value of the mass ratio,
\begin{equation}
    \delta \dot M_\omega(t) = \left| \frac{1}{\sqrt{2 \pi} \sigma} \int e^{- \frac12 (t-t')^2/ \sigma^2} e^{i \omega t'} \dot M(t') dt' \right|,
    \label{eqn:dmdot}
\end{equation}
where $1/\omega \ll \sigma \ll \tau$; $\sigma$ is the window of time over which one wishes to determine the variability, and we typically choose $\sigma = 30$ orbits to calculate this quantity.

$\delta \dot M_{\omega}(t)$ is determined at different specific frequencies $\omega$, and the analytic expression for $q(t)$ is then used to convert this into a function of $q$, $\delta \dot M_{\omega}(q)$.  This is the quantity shown in Figure 6.

For mass ratios below $0.2$, variability is dominated by the orbital and half-orbital timescale.  Above $q = 0.2$, there is a significant variability on a four to five orbit timescale similar to what has been seen by \cite{2012ApJ...749..118S}, \cite{2013MNRAS.436.2997D} and \cite{2014ApJ...783..134F}.  Note, we can demonstrate this variability is not locked to a multiple of the binary orbit; this timescale smoothly moves between four and five orbits as $q$ is varied.  This transition at $q = 0.2$ has also been noted by \cite{2020AAS...23525301N}.   This longer-timescale variability is indicative of episodic accretion from a build-up of mass at the cavity edge, which falls onto the binary all at once, approximately every five orbits.  This is relevant for observations of variable quasars such as PG-1302, for which a binary might explain variability on orbital timescales \citep[as already suggested by][]{2015Natur.525..351D} but this would also predict a longer-timescale variability that is only observable after many orbits \citep{2015MNRAS.452.2540D, Charisi+2015}.


The five-orbit accretion cycle is shown in Figure \ref{fig:saw}.  One can clearly see variability on the half-orbit timescale, in addition to the five-orbit variations, which resemble a sawtooth pattern.  The accretion cycle is quite distinct, and makes specific predictions for binary quasars; one expects a very rapid rise, followed by a slow decay (never the other way around).  This rapid rise also provides a possible explanation for ``changing look" quasars, which shift very rapidly from low-luminosity to a high-luminosity state \citep{2016MNRAS.457..389M, 2019ApJ...874....8M}.  One can model this pattern using the following form:


\begin{equation}
    \dot M(t) = \dot M_0 \left( 1 - B ~\rm tan^{-1}( \rm tan( \Omega_B t / 5 ) ) \right)
\end{equation}
\noindent
where $B = 0.25$ (the form $\rm tan^{-1}(\rm tan(x))$ is merely a convenient way of achieving the sawtooth pattern).  The amplitude $B$ captures the jump in luminosity, which should be proportional to the variability measured on the five-orbit timescale.

This sawtooth pattern could be a possible ``smoking gun" for AGN variability being connected to binary-disk dynamics.  If the system is observed for long enough, one expects to see a sharp rise in the accretion rate every five orbits.  For example, in the case of PG-1302, a speculative possible light curve is plotted in the lower panel of Figure \ref{fig:saw}, accompanied by observations from the Catalina Real-time Transient Survey \citep[CRTS,][]{2015Natur.518...74G} and ASAS-SN \citep{2014ApJ...788...48S, 2017PASP..129j4502K, 2018ApJ...859L..12L}, along with data from photographic plates over the past century, which have been digitized as part of the the Digital Access to a Sky Century at Harvard \citep[DASCH,][]{2009ASPC..410..101G}.  If the four-year variability observed in PG-1302 is indicative of a binary, then we predict a sharp rise in the luminosity by about a magnitude, possibly peaking around the year 2027.  Observations of this quasar over the next decade will be able to determine whether this binary accretion model is accurate. 

Additionally, PG-1302 was just one of many short-period variable quasars that have been discovered, some of which have shorter periods \citep{2015MNRAS.453.1562G, 2016MNRAS.463.2145C, LiuGezari+2019}.  It is possible this sawtooth pattern might be discovered in another example.

\section{Discussion} \label{sec:disc}
This study measured accretion onto each component of a binary system, its accretion variability, and the total torque felt by the binary due to the circumbinary accretion disk.  This information can be used to determine how the binary will evolve as mass is accreted and make predictions for a variety of observed astrophysical binary systems.  For example, as the accretion drives the binary towards equal mass, one can ask how much mass does one need to accrete to make a ``twin binary", which is relevant for observations of binary stars that suggest a significant population of twins with mass ratio $q>0.95$ \citep[e.g.][]{2019arXiv190610128E}.  Additionally, if one assumes all mass after formation was gained by accretion, it may be possible to run back the clock and constrain the initial masses of the two protostars.

\subsection{Accretion}

Given equation (\ref{eqn:lambda}) for the relative accretion rate, it is possible to calculate the mass ratio at a future time.
\noindent
First, define

\begin{equation}
    \lambda \equiv \dot M_2 / \dot M_1
\end{equation}
and assume that it is given by (\ref{eqn:lambda}).  The mass ratio growth rate is then

\begin{equation}
    \dot q = \frac{\dot M_2}{M_1} - \frac{\dot M_1 M_2}{M_1^2} = \frac{\dot M_1}{M_1} ( \lambda - q)
\end{equation}

\begin{equation}
    = \frac{\dot M_B}{M_B} \frac{(1+q) (\lambda(q) - q)}{1 + \lambda(q)},
\end{equation}
where $M_B=M_1+M_2$ is the total binary mass and $\dot M_B = \dot M_1 + \dot M_2$  is the total mass accretion rate onto the system.  It follows that

\begin{equation}
    d{\rm ln}( M_B ) = \frac{(1+\lambda(q)) dq}{(1 + q)( \lambda(q) - q)}.
\end{equation}
\noindent
Integrating this with $\lambda(q)$ given by (\ref{eqn:lambda}) leads to the expression

\begin{equation}
    M_B^{\rm final} / M_B^{\rm init} = w(q_{\rm final})/w(q_{\rm init}),
    \label{eqn:qfin}
\end{equation}
where
\begin{equation}
    w(q) \equiv (1-q)^{-a}(1+q)(1+q/1.1)^{-b},
\end{equation}
and $a = 1/1.9 = 0.53$ and $b = 0.9/1.9 = 0.47$.

Under the simplifying assumptions that initially $M_2 \ll M_1$ and $M_2 \ll M_{\rm acc}$ where $M_{\rm acc}$ is the total accreted mass, 

\begin{equation}
    M_{\rm acc}/M_1^{\rm init} = w(q_{\rm final})-1.
\end{equation}

\noindent
Note, if one defines a ``twin" binary as $q > 0.95$ \citep[as in][]{2019arXiv190610128E} this can be used to show the binary needs to accrete at least 6 times its initial mass to become a twin, if starting from an extreme mass ratio ($w(0.95) \approx 7$).  On the other hand, starting from a more typical mass ratio $q_{\rm init} = 0.5$ necessitates that the binary accrete about three times its initial mass before becoming a twin ($w(0.95)/w(0.5) \approx 4$).  This suggests that observed twin binaries may have been born with a quarter of their present total mass and accreted the rest.

One can perform a similar calculation to determine the primary mass as a function of $q$, leading to the expression

\begin{equation}
    M_1 / M_1^{\rm init} = g(q) / g(q_{\rm init}),
\end{equation}
where
\begin{equation}
    g(q) \equiv (1-q)^{-a}(1+0.9q)^{-b}.
\end{equation}
\noindent
This expression can be used to determine a relationship between $M_1$ and $M_2$ as the binary accretes material from the disk.

A simpler expression can be obtained when the function $g(q)$ is approximated by the following:

\begin{equation}
    g(q) \approx (1-q^m)^{1/m}
\end{equation}
where $m = 1.85$.  This expression is accurate to within $2\%$ everywhere, and therefore just as good as the other expression for $g(q)$, which was obtained via fitting a curve to numerical results.  Then, for brevity define $M_1^{\rm init} \equiv M_0$, and after some algebra one arrives a the expression

\begin{equation}
    M_2^m = M_1^m - M_0^m
\end{equation}
where again $m=1.85$ and $M_0$ is a constant, which can be interpreted as the value of the primary mass when the secondary first forms.  This expression can be used to determine the relationship between $M_1$ and $M_0$ as the binary accretes.  Note that if accretion were divided equally between the primary and secondary (i.e. $\lambda(q) = 1$), then the same expression would work with $m = 1$.

\subsection{Torque}

One can similarly relate the torque to the accretion rate, determining the final separation as a function of the mass accreted.  As a simplification, assume the torque is always equal to $\kappa \dot M j_B$, and assume the binary remains circular.  Note that $\kappa$ as defined in our study should equal $l_0 - q/(1+q)^2$, where $l_0$ is the torque eigenvalue as defined in \cite{2019arXiv191004763M}.  This relationship assumes that angular momentum is accreted at the specific angular momentum of the binary (in agreement with our study and also with the findings of \cite{2020ApJ...891..108D}), so that the accreted angular momentum is $\dot J_{acc} = \dot M j_{bin} = \dot M a^2 \Omega_B q/(1+q)^2$.  Now,

\begin{equation}
    \dot j_B = T/M_B = \kappa \dot M_B a^2 \Omega_B / M_B.
\end{equation}

But $j_B = q/(1+q)^2 a^2 \Omega_B$.  So,

\begin{equation}
    \kappa d{\rm ln} M_B = \frac{q}{(1+q)^2} d{\rm ln} (a^2 \Omega_B).
\end{equation}

For a circular binary, $a^2 \Omega_B = (G M_B a)^{1/2}$.  Then,

\begin{equation}
    \kappa d{\rm ln} M_B = \frac{q}{(1+q)^2} \frac12 ( d{\rm ln} a + d{\rm ln} M_B ).
\end{equation}

Solving for the logarithmic slope:

\begin{equation}
    \frac{ d{\rm ln} a }{ d{\rm ln} M_B } = 2 \kappa \frac{(1+q)^2}{q} - 1.
\end{equation}

This slope is in the range of $+3$ to $+5$ for a wide range of disk parameters.  Call this slope $\delta$:

\begin{equation}
    \delta \equiv \frac{ d{\rm ln} a }{ d{\rm ln} M_B } =  2 \kappa \frac{(1+q)^2}{q} - 1.
\end{equation}

Substituting this formula into equation (\ref{eqn:qfin}) one can determine the change in separation as a function of the change in mass ratio:

\begin{equation}
    \frac{ a_B^{\rm final} }{ a_B^{\rm init} } 
    = 
    \left( \frac{ M_B^{\rm final} }{ M_B^{\rm init} } \right)^{\delta}
    = 
    \left( \frac{ w(q_{\rm final}) }{ w(q_{\rm init}) } \right)^{\delta}.
\end{equation}

For twin binaries we found that $w(q_{\rm final})/w(q_{\rm init})$ could be as small as $4$ for typical $q_{\rm init}$.  This means for twins that have grown in the range where $\delta \approx 4$, the binary can increase its separation by $4^4$ or several orders of magnitude.  This possibly explains the twin binaries seen at very large separations \citep{2019arXiv190610128E} though one would expect the maximum separation to be set by the disk size.

How close can the binary get while $q \ll 1$ and the torque is still negative?  In this limit, one can assume that all the mass is accreted by the secondary, so $d \rm ln M_B = dM_2/M_1 = dq$.  Then

\begin{equation}
    \frac{d \rm ln a}{d \rm ln q} = 2 \kappa.
\end{equation}

Say $\kappa \approx -1$ as shown in the center panel of Figure \ref{fig:over}.  Then

\begin{equation}
    a \propto q^{-2},
\end{equation}
so that one can determine how close the binary gets, as a function of the initial separation and mass ratio:

\begin{equation}
    \frac{a_{\rm closest}}{a_0} = \left( \frac{q_0}{q_{\rm turn}} \right)^2 \approx 
    \left( \frac{q_0}{0.05} \right)^2.
\end{equation}
where $q_{\rm turn} \approx 0.05$ is the mass ratio at which the torque changes sign and the binary executes a U-turn, now moving outward rather than inward.  So, if one starts from $q = 0.01$, the secondary could migrate inward by a factor of $25$ before the U-turn.

So how do we get close binaries, if the torque is always positive (for $q > 0.05$)?  One possibility is given by eccentricity.  For mass ratios greater than $0.2$, the cavity is eccentric and therefore may induce an eccentricity on the binary.  This has already been found by \cite{2019ApJ...871...84M}, who also found a reduced value for $\dot a$, the rate of change of semimajor axis.  It is possible that somewhere in parameter space (either lower viscosity than we probed, or perhaps thinner disks), the disk might drive the binary to a large enough eccentricity to flip the sign of $\dot a$ to be negative again.  So far, this has not been observed in any numerical calculations, but we will explore this idea in a future study.

If eccentricity caused the binary to migrate inwards for mass ratios close to unity, this would make a prediction that the most massive twins should be close binaries, as the binaries at large separations must have stopped accreting shortly after becoming twins.  So far, this idea appears consistent with observations.  For example, see Figure 35 of \cite{2017ApJS..230...15M} or Figure 11 of \cite{2019arXiv190610128E}, which show large separations appear for less massive twins, but that close twins tend to be more massive.

\subsection{Variability}

Accretion variability provides a potentially observable signature of a binary black hole system.  A well-known example is the quasar PG-1302, which appears to exhibit periodic (possibly sinusoidal) variability in the flux as a function of time.  Particularly interesting is that over the last five years, PG-1302 appears to have broken its simple sinusoidal evolution, suggesting that perhaps there is a longer-timescale variability in the source.  This potentially can be connected back to the interaction between the disk and the binary (although there is no direct evidence for a binary, the regular period of the variability generally points to a binary as a possible explanation).  It has already been observed in numerical calculations that accretion can be variable on a five-orbit timescale \citep[and sometimes longer depending on disk properties; see ][]{2014ApJ...783..134F} in addition to the simple variability seen on an orbital timescale.

If this variability is really due to the accretion onto a binary, then we predict a sharp rise in the accretion rate sometime in the next few orbits (sometime between now and 2030).  This would be the signature of the piled-up mass all falling at once onto the binary.  If no such rise is seen, then either this suggests there is no binary at the center, or perhaps the binary has a mass ratio below $0.2$.  Additionally, there are many other black hole binary candidates identified via periodic variability \citep{2015MNRAS.453.1562G, 2016MNRAS.463.2145C, LiuGezari+2019}, with shorter periods that might make it easier to observe this long-term behavior.

Additionally, the long-term sawtooth pattern is quite distinct, always rising rapidly and decaying slowly in a periodic cycle.  This sharp rise might provide a possible explanation for ``changing look" quasars that also  exhibit a rapid rise in magnitude over a very short period of time accompanied by a correspondingly drastic change in the broad emission lines \citep{2016MNRAS.457..389M, 2019ApJ...874....8M}. If the broad emission line region lies at the inner region of the circumbinary disk cavity, as explored in \citet{2015MNRAS.452.2540D}, then the same orbiting feature that causes the long timescale accretion rate variations could also cause extreme broad line variations in changing look quasars. 
Confirming this would require observation over longer timescales, either to capture the short-term variability on the orbital or half-orbit timescales, or to observe the slow decline that should follow the rapid rise.
\\
\acknowledgments

We are grateful to Kareem El-Badry, Selma de Mink, Lars Hernquist, and Magdalena Siwek for helpful comments and discussions.  PCD is supported by Harvard University through the ITC Fellowship.  The computations in this paper were run on the FASRC Odyssey cluster supported by the FAS Division of Science Research Computing Group at Harvard University.

We thank the anonymous referee for the thoughtful review of our work.

\bibliographystyle{apj} 


\end{document}